\documentclass{iaus}
\usepackage{graphicx}
\usepackage{epstopdf}
\usepackage{natbib}

\title[Clues to Radial Migration from the Properties of Outer Disks] 
{Clues to Radial Migration from the Properties of Outer Disks}

\author[Ro\v{s}kar et al.]   
{Rok Ro\v{s}kar$^{1\ast}$,
Victor P. Debattista$^2$,
Thomas R. Quinn$^1$,
Greg S. Stinson$^3$,
James Wadsley$^3$,
Tobias Kaufmann$^4$}

\affiliation{$^1$Astronomy Department, University of Washington, 
Box 351580, Seattle, WA 98195, USA 
\\ $^{\ast}$email: {\tt roskar@astro.washington.edu} \\[\affilskip]
$^2$RCUK Fellow at Centre for Astrophysics, University of Central
Lancashire, Preston, PR1 2HE, UK \\[\affilskip]
$^3$Department of Physics and Astronomy, McMaster University, 
Hamilton, ON, L8S 4M1, Canada \\[\affilskip]
$^4$Department of Physics and Astronomy, Centre for Cosmology,
University of California, Irvine, CA 92697}

\pubyear{2008}
\volume{254}  
\jname{The Galaxy Disk in Cosmological Context}
\editors{J. Andersen, J. Bland-Hawthorn \& B. Nordstr\"{o}m}
\begin{document}

\maketitle

\begin{abstract}
The outer disks of galaxies present a unique laboratory for studying the process
of disk formation. A considerable fraction of observed disks exhibit a break in their
surface brightness profiles. The ubiquity of these features points to a crucial aspect
of disk formation which must be explained. Recent theoretical work suggests 
that such breaks are related to significant amounts of radial migration. We discuss
the current observational evidence which supports this picture. 

\end{abstract}

\firstsection 
\section{Introduction}
The majority of the thin disk forms out of gas quiescently cooling and collapsing inside
the host dark matter halo following the last major merger \citep{Brook:2004}. 
While the inner parts of the stellar thin disk are entangled with the pre-merger material,
the outer parts evolve in relative solitude, modulo interactions with substructure components present
within the host halo. The outer parts of galactic disks therefore provide us with a direct
view of disk assembly in progress. 

Since \citet{van-der-Kruit:1979,van-der-Kruit:1987} it has been 
known that the surface brightness profiles of disk galaxies may not always 
follow a simple single-exponential law. 
In a recent work using data from the Sloan Digital Sky Survey (SDSS),
\citet{Pohlen:2006} showed that $\sim60\%$ 
of nearby disk galaxies have downward-bending
surface brightness breaks, traditionally termed ``truncations''. Disk breaks have also
been observed in the distant universe out to a redshift of z$\sim$1 
\citep{Perez:2004,Trujillo:2005,Azzollini:2008a}, further implying that they 
are a generic feature of disk formation. 

Several theories for the formation of downward-bending breaks have been suggested. The
most common interpretations include angular momentum-limited collapse 
\citep{van-der-Kruit:1987, van-den-Bosch:2001},
star formation threshold either due to a critical gas density \citep{Kennicutt:1989} or a 
lack of a cool equilibrium ISM phase \citep{elmegreen:1994,Schaye:2004}. Alternatively, breaks have also been attributed to angular momentum redistribution \citep{Debattista:2006, Foyle:2008}.

\section{Connection Between Radial Migration and Outer Disks}
Recently, \citet{Roskar:2008} (hereafter R08) investigated the formation of outer disk 
breaks via $N$-body + Smooth Particle Hydrodynamics (SPH) simulations of an isolated 
galaxy forming through dissipational collapse. In the simulation, a break in the stellar 
mass density profile is seeded by a drop in the star formation rate, which is associated
with a rapid drop in the gas surface density. However, the stellar profile extends by several
kpc beyond the outermost star forming radius -- the outer exponential is populated almost exclusively
by stars that migrated there through resonant interactions with spiral arms. 
Large migrations, which preserve the circularity of stellar orbits, are possible when 
stars are scattered by the corotation resonance of transient spirals \citep{Sellwood:2002}.

\begin{figure}
	\centering
	\includegraphics[width=4in]{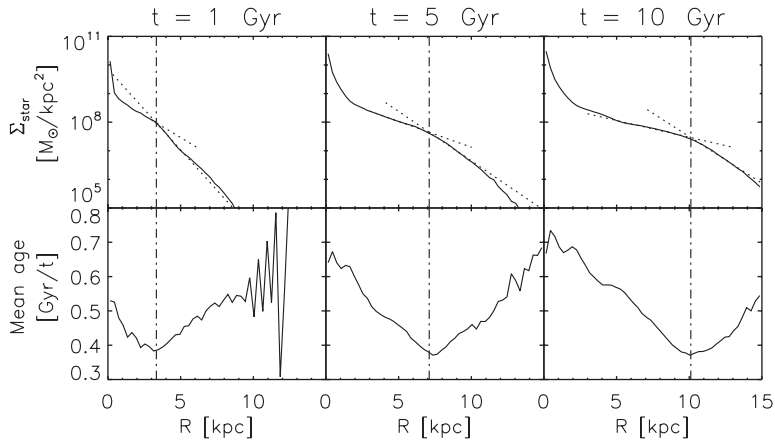}
	\caption{The stellar surface density (top panels) and mean age profiles (bottom panels) 
		at three different times in the simulation. The dashed lines in the upper panels are 
		double exponential fits to the surface density profile. The vertical dashed lines indicate
		the location of the break in all panels. At all times, the break corresponds to a
		minimum in the stellar mean age, which arises due to the significant radial migration
		of stars.}
	\label{fig:dens_plots}
\end{figure}

The disk breaks simulated in R08 are therefore 
a consequence of the interplay between 
a drop in the star formation rate and secular evolution driven primarily by recurring transient spirals. 
Figure~\ref{fig:dens_plots} shows the stellar surface density 
(top panels) and the associated mean age profiles
(bottom panels) at three different times in the simulation. The break in the stellar
surface density is associated with an inflection in the mean age profile, which is a direct 
result of substantial radial redistribution of stellar material. The outer disk thus offers an
opportunity to gain insight into the radial migration process, which may significantly 
affect the entire disk. 

\section{Observations of Disk Breaks in Support of Radial Migration}

Radial migration irreversibly alters the properties of the underlying disk stellar populations. 
Disentangling the signatures of radial migration in the inner regions of disks, which 
have been heavily influenced by a myriad of other processes, is nontrivial. It is therefore 
imperative to attempt to learn about radial migration from the outer disks. 

A first clue that the evolution presented in R08 may be common is found 
in several of the galaxies presented in \citet{Wevers:1986}, where the drop 
in gas surface density corresponds to a break in the surface brightness profile and
a minimum in the color profile (see, for example, the profiles for NGC~3726, NGC~4242, 
and NGC~5371). 
Using multi-band photometry and stellar population synthesis models, \citet{MacArthur:2004} 
presented radial stellar population trends for a sample of nearby galaxies. From 
their Figure 20, it is apparent that age minima are common in the radial 
age profiles for the galaxies in their sample, though
it is unclear whether those minima correspond to breaks in the stellar surface brightness. The 
minima do, in general, appear around 2~-~2.5 disk scale-lengths, which is also roughly the 
mean break scale radius observed in \citet{Pohlen:2006}. This suggests that
the age minima are indeed associated with profile breaks.
A positive age gradient has also been observed beyond the break of M33 via 
CMD modeling of resolved stellar populations, 
but this result is also inconclusive because no age information exists for the inner disk 
regions \citep{Barker:2007a}. In NGC~7793, the GHOSTS team has observed a flattening
of the radial star-count profile beyond the break with increasing age of stellar population, which
implies an increasing mean stellar age (de Jong \& Radburn-Smith, \textit{private communication}).
However, as with the age profile of M33, there is currently no information about the age gradient
interior to the break of NGC~7793. 

\begin{figure}
	\centering
	\includegraphics[width=5in]{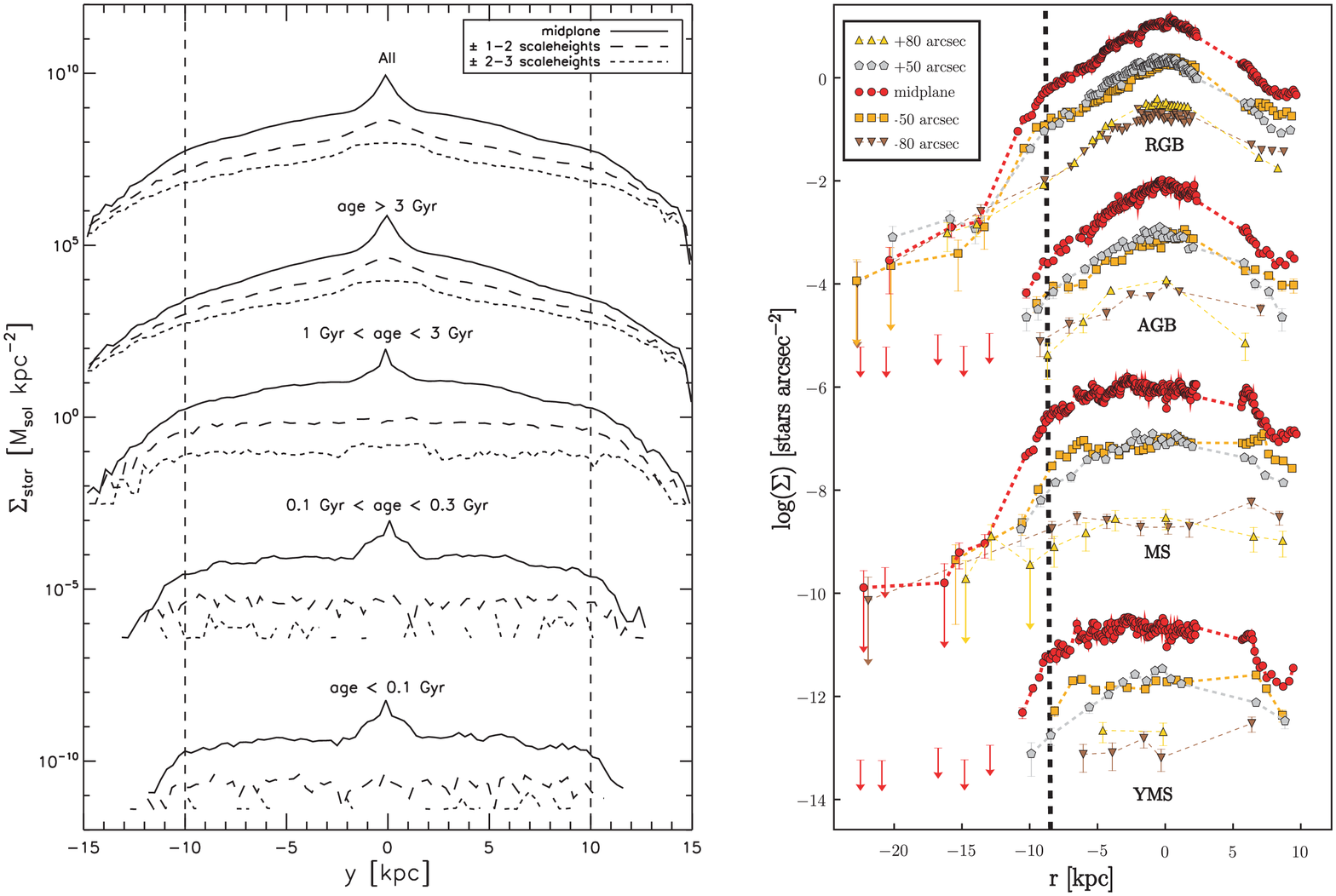}
	\caption{Edge on stellar density profiles from the simulation of \citet{Roskar:2008} (left). The
		break is in the same location regardless of the age of stellar population and 
		height above the plane, consistent with the observations of NGC~4244 
		\citep{de-Jong:2007} (right).}
	\label{fig:edgeon}
\end{figure}

In the case of the edge-on galaxy NGC~4244, observations of resolved stars 
reveal that the break in the stellar profile
occurs in the same place for stars of all ages \cite{de-Jong:2007}. As shown in 
Figure~\ref{fig:edgeon}, models presented in R08 naturally give rise to this 
phenomenon. In the absence of migration, the observed stellar populations of NGC~4244
imply that either the star formation threshold has been in the same place for $\sim$10 Gyr, 
which seems unlikely given that disks are believed to form inside-out 
\citep{Munoz-Mateos:2007}.
Alternatively, the break could have formed on timescales $< 100$~Myr in order
to produce the observed feature (either by dynamics or sudden gas expulsion), 
which seems highly unlikely given that typical dynamical 
times in those parts of galactic disks are considerably longer. Efficient radial 
migration coupled with a star formation threshold therefore provides a very 
plausible disk break formation mechanism for this particular galaxy. 

\citet{Azzollini:2008} presented a detailed study of color gradients in a sample of disk 
galaxies from the GOODS-South field. Their sample includes galaxies in the redshift 
range $0.1 < z < 1.1$. They find that regardless of redshift, the galaxies with observed
downward-bending breaks reveal a minimum in colors roughly analogous to 
rest-frame $u - g$. Importantly, the color profiles for the anti-truncated disks in their sample
show no such feature, arguing against the possibility of a systematic observational 
error being responsible for the signature. It is possible that the positive color gradient beyond 
the break is caused by a change in the metallicity gradient. This seems unlikely 
given that the local sample of \citet{MacArthur:2004} shows no pronounced 
inflections in the metallicity, while age minima are certainly present. 
According to the authors, the most plausible explanation 
for the reversal of the color gradient in \citet{Azzollini:2008} is that the age of the underlying stellar 
population increases beyond the break, in line with the theoretical predictions of R08. 

In summary, these various observational results strongly imply that radial
migration is indeed an important effect at all stages of spiral disk evolution. 

\section{Conclusions}

We have argued that outer disk breaks are a phenomenon that is not only common 
in observed systems, but that its mechanism of formation may provide important 
insights into disk evolution as a whole. Substantial observational evidence taken in the 
context of recent theoretical models suggests that at least a fraction of outer parts of late-type galactic 
disks form due to substantial radial migration of fully-formed stellar material. Such evolution
affects these extreme outer regions of galaxies, and profoundly 
impacts the properties of stellar populations in the entire disk.

\bibliographystyle{apj}

\end{document}